\documentclass[10pt,twocolumn,cleanfoot]{config/asme2e}

\usepackage{empheq, amssymb}

\usepackage{lmodern}
\usepackage{amsmath}
\usepackage{amsfonts}
\usepackage{amssymb}

\usepackage{bm} % bold for greek symbols

\usepackage{txfonts}

\usepackage[T1]{fontenc}
\usepackage[utf8]{inputenc} % accept different input encodings

\usepackage[dvipsnames]{xcolor} % defines colors

\usepackage[american]{babel} %  culturally-determined typographical rules

\usepackage{graphicx}

\usepackage{epstopdf}

\usepackage[noadjust]{cite} % grouped citations

\usepackage{ltxcmds}

\usepackage{mathtools}

\usepackage{subcaption}% for subfigures

\usepackage{multirow} %for tables 

\usepackage{blindtext}

\usepackage{hyperref}

\usepackage{tikz} % for drawings
\usetikzlibrary{shapes.geometric, arrows, topaths, shapes,decorations.text, positioning,shadows,trees,shapes.arrows, fadings,arrows.meta,graphs,graphs.standard}
\usetikzlibrary{fit,calc,backgrounds}

\usepackage{enumitem}

\usepackage{ifthen}

\usepackage{etoolbox}

\usepackage{cancel}

\usepackage{empheq}

\usepackage{fontawesome}

\usepackage{soul}

\usepackage{flushend} % balance columns on last page

\usepackage{textcomp}

\usepackage{lipsum}

\definecolor{light-gray}{gray}{0.4}
\definecolor{box-gray}{gray}{1}

\hypersetup{
    unicode=false,          % non-Latin characters in Acrobat’s bookmarks
    pdftoolbar=true,        % show Acrobat’s toolbar?
    pdfmenubar=true,        % show Acrobat’s menu?
    pdffitwindow=false,     % window fit to page when opened
    pdfstartview={FitV},    % fits the width of the page to the window
    pdftitle={A Framework and Prototype for a Navigable Map of Datasets in Engineering Design and Systems Engineering},    % title
    pdfauthor={H. Sinan Bank and Daniel R. Herber},     % author
    pdfsubject={},   % subject of the document
    pdfkeywords = {}, % list of keywords
    pdfnewwindow=true,      % links in new window
    colorlinks=true,
    allcolors=blue
}

\usepackage{nomencl}

\renewcommand\nomgroup[1]{%
  \item[\bfseries
  \ifstrequal{#1}{V}{ Variables}{%
  \ifstrequal{#1}{B}{ Subscripts}{%
  \ifstrequal{#1}{P}{ Notation}{%
  \ifstrequal{#1}{A}{ Acronyms}{}}}}]
}

\makenomenclature



\usepackage{dblfloatfix}
\usepackage{float}
\usepackage{placeins} % for \FloatBarrier
\definecolor{block-gray}{gray}{0.95}

\usepackage[framemethod=TikZ]{mdframed}

\usepackage{xpatch}

\makeatletter
\xpatchcmd{\endmdframed}
  {\aftergroup\endmdf@trivlist\color@endgroup}
  {\endmdf@trivlist\color@endgroup\@doendpe}
  {}{}
\makeatother

\usepackage{fontawesome}

\usepackage{titlesec}

\usepackage{hyperref}

\newcommand{\doi}[1]{{doi:~\href{http://doi.org/#1}{#1}}\rmFullStop}

\newcommand*{\rmFullStop}{\rmifnextchar{.}{}{}}

\makeatletter
\newcommand{\rmifnextchar}[3]{%
  \begingroup
  \ltx@LocToksA{\endgroup#2}%
  \ltx@LocToksB{\endgroup#3}%
  \ltx@ifnextchar{#1}{%
    \def\next{\the\ltx@LocToksA}%
    \afterassignment\next
    \let\scratch= %
  }{%
    \the\ltx@LocToksB
  }%
}
\makeatother

\usepackage{colortbl}

\definecolor{light-gray}{gray}{0.6}

\newcommand{\xsection}[1]{\section[#1]{\MakeUppercase{#1}}}

\usepackage{algorithm2e}
\usepackage{xcolor}

\SetAlFnt{\footnotesize}

\SetAlCapSty{xAlCapSty}

\definecolor{commentcolor}{HTML}{1E4D2B}

\SetCommentSty{xCommentSty}

\SetNlSty{mynlfont}{}{} 

\LinesNumbered

\SetSideCommentRight

\DontPrintSemicolon

\RestyleAlgo{algoruled}

\usepackage{algorithm2e}
\usepackage{etoolbox}
\usepackage{xstring}
\usepackage{calc}
 
\newlength{\xalgowidth}
\newlength{\xalgoremainder}
\newlength{\xindentwidth}

\newenvironment{vAlgorithm*}[3][]{% before
  \setlength{\xalgowidth}{#2} % set algorithm width from second input
  \setlength{\xindentwidth}{#3} % set indent width from third input
  \setlength{\xalgoremainder}{\textwidth-\xalgowidth} % calculate indent to center the float
  \SetCustomAlgoRuledWidth{\xalgowidth} % set the rule width
  \IncMargin{\xindentwidth}
  \begin{algorithm*}[#1]
}% end before
{% after
  \end{algorithm*} 
  \DecMargin{\xindentwidth}
}% end after

{% after
  \end{algorithm} % column
  \DecMargin{\xindentwidth}
}% end after

\makeatletter
\patchcmd{\@algocf@start}{%
\begin{lrbox}{\algocf@algobox}%
}{%
\rule{0.5\xalgoremainder}{\z@}% indent
\begin{lrbox}{\algocf@algobox}%
\begin{minipage}{\xalgowidth}%
}{}{}
\patchcmd{\@algocf@finish}{%
\end{lrbox}%
}{%
\end{minipage}%
\end{lrbox}%
}{}{}
\makeatother

\usepackage{accents}

\definecolor{needcolor}{HTML}{C62828}

\confshortname{}
\conffullname{}
\confdate{}
\confmonth{}
\confyear{}
\confcity{}
\confcountry{}
\papernum{}

\title{A Framework and Prototype for a Navigable Map of Datasets in Engineering Design and Systems Engineering}

\author{H.~Sinan~Bank\thanks{Corresponding author: \href{mailto:sinan.bank@colostate.edu}{sinan.bank@colostate.edu}}
\affiliation{
Department of Systems Engineering \\
Colorado State University \\
Fort Collins, CO 80523 \\
}
}
\author{Daniel~R.~Herber
\affiliation{
Department of Systems Engineering \\
Colorado State University \\
Fort Collins, CO 80523 \\
}
}

\usepackage{amsmath}
\usepackage{graphicx}

\usepackage{microtype}

\makeatletter
\def\@maketitle{%
  \newpage
  \null\vspace*{-26pt}%
   \vbox{\hbox to \textwidth{\begin{tabular}{@{}c@{\hskip3pc}}
   \hbox to 46pt{\vbox to 46pt{\vss\hsize46pt\vss}}\end{tabular}\hss
  \vbox{\hsize37pc\scriptsize\sf\vskip\baselineskip%
  \bannerfnt\begin{flushright}%
  \ifx\@conffullname\empty\else\hfill \@conffullname\par\fi
  \ifx\@confshortname\empty\else\hfill \@confshortname\par\fi
  \vskip.036in
  \ifx\@confmonth\empty\else\hfill \@confmonth\ \@confdate, \@confyear, \@confcity, \@confcountry\fi
  \vskip.5in
  \ifx\@papernum\empty\else\hfill {\pnumfnt\@papernum}\fi\end{flushright}}
  \hskip1pc}}
  \vskip.15in%
  \vskip .25pc{\large\twlsfb
               \begin{center}\leftskip.5in plus1fill\rightskip\leftskip
               \@title\par\end{center}}
  \vskip2pc{\begin{center}\@author\par\end{center}}\vskip12pt}
\makeatother

\begin{document}
\setlength{\parskip}{0pt}
\setlength{\parsep}{0pt}
\setlength{\headsep}{0pt}
\setlength{\topsep}{0pt}

% equations
\abovedisplayshortskip=3pt
\belowdisplayshortskip=3pt
\abovedisplayskip=3pt
\belowdisplayskip=3pt

\titlespacing*{\section}{0pt}{18pt plus 1pt minus 1pt}{3pt plus 0.5pt minus 0.5pt}

\titlespacing*{\subsection}{0pt}{9pt plus 1pt minus 0.5pt}{1pt plus 0.5pt minus 0.5pt}

\titlespacing*{\subsubsection}{0pt}{9pt plus 1pt minus 0.5pt}{1pt plus 0.5pt minus 0.5pt}

\microtypesetup{nopatch=item}
\maketitle
\microtypesetup{patch=item}

%---------------
\begin{abstract}\noindent
\textit{The proliferation of data across the system lifecycle presents both a significant opportunity and a challenge for Engineering Design and Systems Engineering (EDSE). While this ``digital thread'' has the potential to drive innovation, the fragmented and inaccessible nature of existing datasets hinders method validation, limits reproducibility, and slows research progress. Unlike fields such as computer vision and natural language processing, which benefit from established benchmark ecosystems, engineering design research often relies on small, proprietary, or ad-hoc datasets. This paper addresses this challenge by proposing a systematic framework for a ``Map of Datasets in EDSE.'' The framework is built upon a multi-dimensional taxonomy designed to classify engineering datasets by domain, lifecycle stage, data type, and format, enabling faceted discovery. An architecture for an interactive discovery tool is detailed and demonstrated through a working prototype, employing a knowledge graph data model to capture rich semantic relationships between datasets, tools, and publications. An analysis of the current data landscape reveals underrepresented areas (``data deserts'') in early-stage design and system architecture, as well as relatively well-represented areas (``data oases'') in predictive maintenance and autonomous systems. The paper identifies key challenges in curation and sustainability and proposes mitigation strategies, laying the groundwork for a dynamic, community-driven resource to accelerate data-centric engineering research.}
\end{abstract}

\vspace{1ex}

%---------------
% Section 1: Introduction
\xsection{Introduction}\label{sec:introduction}

%---------------
% Figure 1: Combined Framework Overview + Taxonomy (full-width)
\begin{figure*}[!t]
\centering
\includegraphics[width=0.85\textwidth]{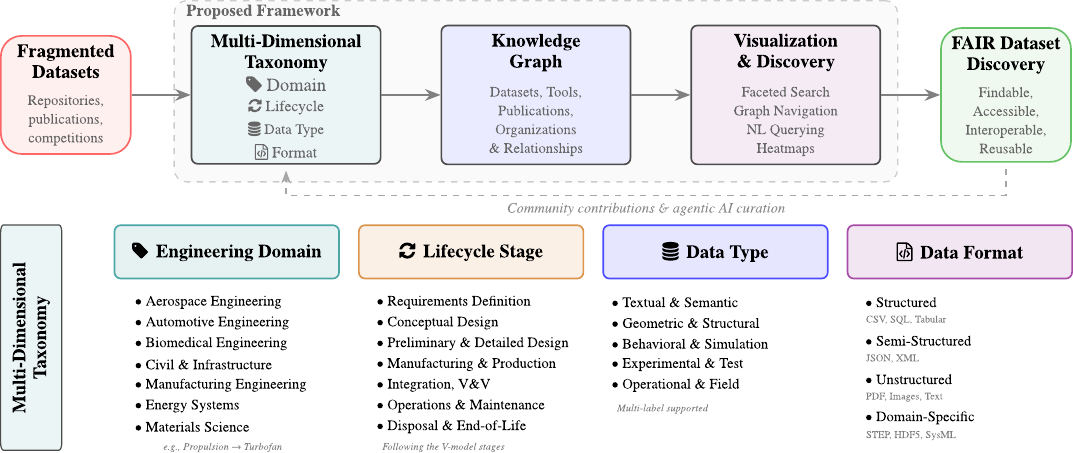}
\caption{Proposed framework (top) and multi-dimensional taxonomy (bottom) for classifying and discovering engineering datasets.}
\label{fig:framework_overview}
\label{fig:taxonomy}
\end{figure*}

The transition to a data-centric paradigm is reshaping engineering disciplines. In Engineering Design and Systems Engineering (EDSE), this shift has produced a significant volume of data spanning the entire system lifecycle, from requirements definition and conceptual design to manufacturing, operations, and disposal. This digital thread offers the potential to enable AI-driven design, optimize complex systems, and validate new engineering methodologies.

However, a primary bottleneck prevents the full realization of this potential: the fragmented, siloed, and often inaccessible nature of engineering datasets. Researchers and practitioners face significant difficulty in discovering, comparing, and reusing relevant data~\cite{Ahmed2025,Felten2025}. Unlike fields such as computer vision and natural language processing, which benefit from large-scale benchmark datasets, engineering design research often proceeds with small, proprietary, or ad-hoc datasets that limit reproducibility and generalizability.

Research has identified a fundamental theory-practice gap that complicates AI system engineering across data quality assurance, model building, and deployment~\cite{Fischer2020}. Data-driven design in the early phases of physical product development faces particular challenges~\cite{Briard2023}. The reliance on data availability represents a significant hindrance, often leading to underexplored design spaces~\cite{Rad2025}. Both Model-Based Systems Engineering (MBSE) and AI face challenges regarding successful deployment, with a key bottleneck being the lack of a coherent foundation to integrate data-driven methods~\cite{Chami2022}. Additional concerns include the reliability and interpretability of AI-driven results, particularly with ``black-box'' models that produce outputs difficult to validate~\cite{Rudin2019}, and barriers related to human trust in AI predictions despite achieving comparable accuracy to human experts~\cite{Schmidt2020}.

This challenge, frequently highlighted within the research community, directly impedes scientific progress. To address this critical need, this paper presents the design of a ``Map of Datasets in EDSE''. The objective is to move beyond a static list of resources and define a dynamic, navigable ecosystem that makes engineering data Findable, Accessible, Interoperable, and Reusable (FAIR). An initial prototype is presented to demonstrate feasibility, while complete implementation and validation of this design are identified as future work.

The contribution of this work is a concrete design comprising three main components, as illustrated in Fig.~\ref{fig:framework_overview} (top):
\begin{enumerate}
    \item A \textbf{multi-dimensional taxonomy} for structured classification of engineering datasets across four dimensions: engineering domain, system lifecycle stage, data type/modality, and data format
    \item A \textbf{knowledge graph architecture} specifying node types, relationship types, and interaction modes for an interactive discovery tool that captures rich semantic relationships between datasets, tools, publications, and taxonomy terms
    \item An \textbf{analysis of the current dataset landscape} to identify gaps (``data deserts'') and relatively well-represented areas (``data oases''), with discussion of mitigation strategies including synthetic data
\end{enumerate}

The remainder of this paper is organized as follows: Section~\ref{sec:background} reviews background and related work on FAIR principles, existing dataset ecosystems, and knowledge graph technologies. Section~\ref{sec:methodology} presents the proposed design, including the multi-dimensional taxonomy and knowledge graph architecture. Section~\ref{sec:results} applies the framework to analyze the current landscape and presents an exemplar dataset catalog. Section~\ref{sec:conclusion} concludes with future work toward an intelligent, agent-driven ecosystem.

%---------------
% Section 2: Background and Related Work
\xsection{Background and Related Work}\label{sec:background}

The problem of data discoverability is a well-recognized challenge across scientific domains, leading to the development of the FAIR (Findability, Accessibility, Interoperability, and Reusability) Guiding Principles for scientific data management~\cite{Wilkinson2016}. While general-purpose data catalogs such as data.gov and Zenodo exist, they often lack the domain-specific structure required to navigate the complexities of engineering data. The engineering community has produced numerous high-value datasets, but they are scattered across institutional repositories, competition websites, and individual publications.

\subsection{Existing Dataset Ecosystems}

Benchmark datasets in prognostics and health management (PHM), such as the NASA C-MAPSS turbofan engine data~\cite{Saxena2008} and the CWRU Bearing Data~\cite{CWRU}, have become foundational for validating new algorithms. Similarly, the computer vision community has demonstrated the power of shared data through benchmarks such as KITTI for autonomous driving~\cite{Geiger2012}. In materials science, the Materials Project has revolutionized discovery by providing a massive, open database of computed properties~\cite{MaterialsProject}.

However, these successes are isolated. There is no unifying framework to connect a dataset of manufacturing line performance from Bosch with a repository of public requirements documents or a collection of CAD models. This lack of a central, structured map makes cross-domain discovery difficult and hinders the development of holistic, lifecycle-aware engineering AI.

\subsection{FAIR Principles and Implementation Barriers}

The FAIR principles provide a framework for managing heterogeneous datasets across the product lifecycle. However, assessments in domains such as lifecycle assessment reveal that although awareness of FAIR data sharing is increasing, implementing specific FAIR guidelines is rarely observed in practice~\cite{Ghose2024}. The challenge of independent and siloed datasets limits transparency and interoperability. Recent work on metadata repositories using Linked Data principles demonstrates approaches for real-time access to heterogeneous source systems~\cite{Eickhoff2020}, while Digital Twin frameworks have been proposed to promote FAIR principles in automotive use-cases~\cite{Hamlaoui2025}.

Poor interoperability among computer-aided engineering (CAE) software tools costs the industry billions of dollars~\cite{Szykman2000}. Fundamental interoperability barriers (e.g., in Systems of Systems) include heterogeneous data and disparate APIs~\cite{Sadeghi2024}. Industrial scenarios present largely unexplored applicability of FAIR principles compared to research data~\cite{Bodenbenner2021}. The aspirational nature of FAIR principles means they do not provide precise guidance for direct implementation into specific domains, complicating practical adoption~\cite{Karakoltzidis2024}.

\subsection{Ontologies and Taxonomies for Engineering Artifacts}

Prior work has established the importance of formal ontologies for organizing engineering design knowledge. Ontologies of engineering artifacts contribute to design knowledge modeling by providing structured taxonomies with rich relationships~\cite{Kitamura2006}. Foundational ontologies such as BFO, GFO, and DOLCE offer frameworks for characterizing and classifying artifacts~\cite{Borgo2009}. Knowledge-Based Engineering Data Management (KBEDM) approaches organize knowledge-intensive activities in modern design processes~\cite{Mario2013}.

Domain-specific ontologies have been developed for CAD/CAE integration, including simulation intent ontologies that formalize analysis parameters and idealization decisions~\cite{Boussuge2019}. Layered ontology architectures---comprising general domain, domain-specific, and application-specific ontologies---represent engineering design knowledge at multiple abstraction levels~\cite{Zhu2009,Zhan2010}. Such approaches enable semantic-level information exchange between heterogeneous applications across the product lifecycle~\cite{Zhan2007}. Ontological engineering has also addressed integration of CAD and GIS for infrastructure management~\cite{Peachavanish2006}. Prior work on representation frameworks for engineering design provides classification based on vocabulary, structure, expression, purpose, and abstraction~\cite{Summers2004}. Efforts to synthesize product knowledge across the lifecycle using upper-tiered ontologies~\cite{Witherell2013} inform the taxonomy proposed in this work.

\subsection{Knowledge Graphs for Data Integration}

Knowledge graphs offer advantages over relational databases for managing diverse, interconnected datasets. Knowledge graphs integrate heterogeneous data from various sources---unstructured, semi-structured, and structured---in a semantically rich way~\cite{Hofer2024}. Graph databases are valuable for data integration because their unfixed structure allows flexibility, unlike relational databases that depend on rigid schemas~\cite{DiPierro2025}. Comparative studies show graph databases significantly outperform relational databases in search query response time~\cite{Lorincz2020}, and are particularly effective for handling large-scale data requiring semantic association and visualization~\cite{Zhang2025}.

The LinkClimate platform demonstrates knowledge graphs for integrating multi-source heterogeneous data with improved interoperability through ontologies~\cite{Wu2022}. Data-centric system design using Knowledge Graphs and Semantic Web technologies provides a framework for data interoperability~\cite{Rojas2021}. The flexibility of graph-based data models means adding new data sources requires significantly less effort than altering relational schemas. Schema-free graph databases can be high-performance replacements for relational databases when handling highly interconnected data~\cite{Kalayci2021}. Graph-based approaches have enabled practical applications such as railway infrastructure systems that merged disconnected relational databases into unified knowledge graphs~\cite{Toledo2025}. This makes knowledge graphs particularly suitable for representing the complex relationships between datasets, tools, methods, and publications in engineering design.

%---------------
% Section 3: Methodology
\xsection{Methodology: A Framework for the EDSE Dataset Map}\label{sec:methodology}

This section presents the proposed framework for a navigable ``Map of Datasets in EDSE''. The design comprises two main components: a multi-dimensional taxonomy for structured classification and a knowledge graph architecture for an interactive discovery tool.

\subsection{The Multi-Dimensional Taxonomy}

A foundational element of the proposed map is a multi-dimensional taxonomy that enables faceted search and flexible organization, as illustrated in Fig.~\ref{fig:taxonomy} (bottom). Unlike traditional flat lists or single-category classification schemes, a multi-dimensional taxonomy allows datasets to be characterized along multiple independent dimensions simultaneously. This approach supports diverse user needs: a researcher seeking time-series data for prognostics can filter by data type, while another seeking aerospace applications can filter by domain, and both can combine criteria to narrow results. Four key \textit{dimensions} are proposed, each structured hierarchically to support navigation from broad categories to specific sub-disciplines.

\subsubsection{Dimension~1:~Engineering Domain and Application Area.} The first dimension categorizes datasets by their primary field of engineering application. This reflects the reality that engineering research and practice are organized around domain-specific communities, conferences, and publication venues. Researchers typically begin their search within their home domain before considering cross-domain resources. The hierarchical structure of this dimension allows users to navigate from broad domains such as Aerospace Engineering, Automotive Engineering, Biomedical Engineering, Civil and Infrastructure Engineering, Manufacturing Engineering, and Energy Systems, down to increasingly specific sub-disciplines. For example, a user might navigate from Aerospace to Propulsion Systems to Turbofan Engines to find the NASA C-MAPSS dataset, or from Automotive to Autonomous Systems to Perception to discover the KITTI benchmark. This hierarchical organization also reveals structural similarities across domains---condition monitoring datasets in aerospace propulsion share methodological characteristics with those in wind turbine drivetrains, even though they originate from different engineering communities.

\subsubsection{Dimension~2:~System Lifecycle Stage.}~The second dimension classifies datasets according to the phase of the systems engineering lifecycle in which they are generated or primarily applicable. This dimension is particularly important for engineering design research because data characteristics, availability, and quality vary dramatically across lifecycle stages. The lifecycle stages follow the general structure of the systems engineering V-model, spanning from early conceptual phases through detailed design, manufacturing, operations, and eventual disposal. The stages include System Requirements Definition, which encompasses stakeholder needs, functional requirements, and specifications; Conceptual Design and Trade Studies, covering architecture exploration and concept evaluation; Preliminary and Detailed Design, addressing component specifications, CAD models, and simulation results; Manufacturing and Production, including process data, quality control, and assembly information; Integration, Verification, and Validation, encompassing test data and certification evidence; Operations and Maintenance, covering field data, sensor streams, and maintenance records; and finally Disposal and End-of-Life, addressing decommissioning and material recovery data. This lifecycle perspective enables users to find data relevant to their specific engineering activity and reveals the digital thread connecting data across stages.

\subsubsection{Dimension~3:~Data Type and Modality.}~The third dimension describes the fundamental nature of the information contained within a dataset, which determines the analytical methods and tools required for its use. Engineering datasets span a remarkable diversity of modalities, from natural language text to three-dimensional geometry to high-frequency sensor signals. Understanding this diversity is essential for researchers seeking data compatible with their methods. The taxonomy distinguishes five primary data types. Textual and Semantic data includes requirements documents, specifications, design rationale, technical reports, and other natural language content that encodes engineering knowledge in human-readable form; these datasets are amenable to natural language processing and information extraction techniques. Geometric and Structural data encompasses CAD models, meshes, point clouds, and other representations of physical form; these require specialized geometric processing algorithms and domain-specific file formats. Behavioral and Simulation data include time-series outputs from physics-based simulations, system models, and digital twins that capture how systems evolve over time; these datasets enable the development and validation of surrogate models and reduced-order approximations. Experimental and Test data comprises measurements from laboratory experiments, component tests, and controlled evaluations that provide ground truth for model validation. Finally, Operational and Field data include sensor streams, maintenance logs, and performance records from deployed systems operating in real-world conditions; these datasets are essential for prognostics, health management, and operational optimization but often come with challenges of noise, missing data, and proprietary restrictions. Importantly, a single dataset may span multiple data types---a digital twin dataset might include both geometric models and behavioral simulations---and the taxonomy supports multi-labeling to capture this richness.

\subsubsection{Dimension~4:~Data Format and Structure.}~The fourth dimension classifies datasets based on their file format and internal organization, which dictates the technical requirements for access and processing. This practical dimension is often overlooked in conceptual discussions but is critically important for researchers who must actually work with the data. The taxonomy distinguishes four categories of data structure. Structured data follows a predefined schema with explicit relationships, including tabular formats such as CSV and relational databases; these are readily ingested by standard data science tools and machine learning pipelines. Semi-Structured data has organizational properties but without rigid schemas, including hierarchical formats such as JSON and XML; these require parsing but offer flexibility for complex nested relationships. Unstructured data lacks predefined organization, including PDF documents, images, and raw text; these require content extraction and interpretation before analysis. Finally, Domain-Specific Formats are specialized representations developed for particular engineering applications, such as STEP and IGES for CAD geometry, HDF5 for large scientific datasets, and SysML/XMI for system models; these require specialized software and domain expertise but preserve rich semantic information. By classifying format alongside content, the taxonomy helps users identify datasets they can realistically work with given their available tools and expertise.

\subsection{Knowledge Graph Architecture for the Navigable Map}

The multi-dimensional taxonomy provides a classification scheme, but realizing a truly navigable map requires an underlying data model and user interface that support flexible exploration. This section describes the proposed architecture for an interactive discovery tool.

\subsubsection{Data Model:~Knowledge Graph.}~A knowledge graph is proposed as the underlying data model, chosen over traditional relational databases for several compelling reasons. First, the graph model offers superior flexibility by representing datasets, taxonomy terms, tools, methods, publications, and organizations as nodes connected by typed relationships (edges). This structure natively supports the faceted classification scheme: a dataset node can connect to multiple taxonomy term nodes across all four dimensions simultaneously, without the awkward join tables required in relational schemas. Second, knowledge graphs enable rich semantic relationships beyond simple classification. Relationships such as: 
\begin{itemize}[nosep,label=$\circ$]
\item \texttt{dataset\_A \ used\_in \ publication\_X}
\item \texttt{tool\_Y \ compatible\_with  \ format\_Z}\item \texttt{dataset\_A  \ derived\_from  \ dataset\_B}
\item \texttt{method\_M  \ validated\_on  \ dataset\_D}
\end{itemize}

\noindent can be explicitly represented and queried. These relationships capture the intellectual connections within the research community that are invaluable for discovery but impossible to represent in flat catalogs. Third, knowledge graphs support extensibility: new entity types, relationship types, and attributes can be added incrementally without schema migrations that would disrupt existing queries. This is essential for a community resource that must evolve with the field. Fourth, adoption of standard vocabularies such as DCAT (Data Catalog Vocabulary) ensures interoperability with other data catalogs and the broader semantic web ecosystem.

The knowledge graph schema includes five primary node types. Dataset nodes carry metadata including title, description, source URL, license, DOI, size, temporal coverage, and quality metrics. TaxonomyTerm nodes represent each term in the four taxonomic dimensions, connected by hierarchical \texttt{parent\_of} relationships that enable roll-up queries and faceted navigation. Publication nodes represent papers, reports, and other scholarly works that use or describe datasets. Tool nodes represent software packages and libraries compatible with specific data formats or analysis types. Organization nodes represent institutions that create, host, or maintain datasets.

%===============================================================================
% FIGURE: Prototype Screenshots (full page width)
%===============================================================================
\begin{figure*}[!t]
\centering
\includegraphics[width=\textwidth]{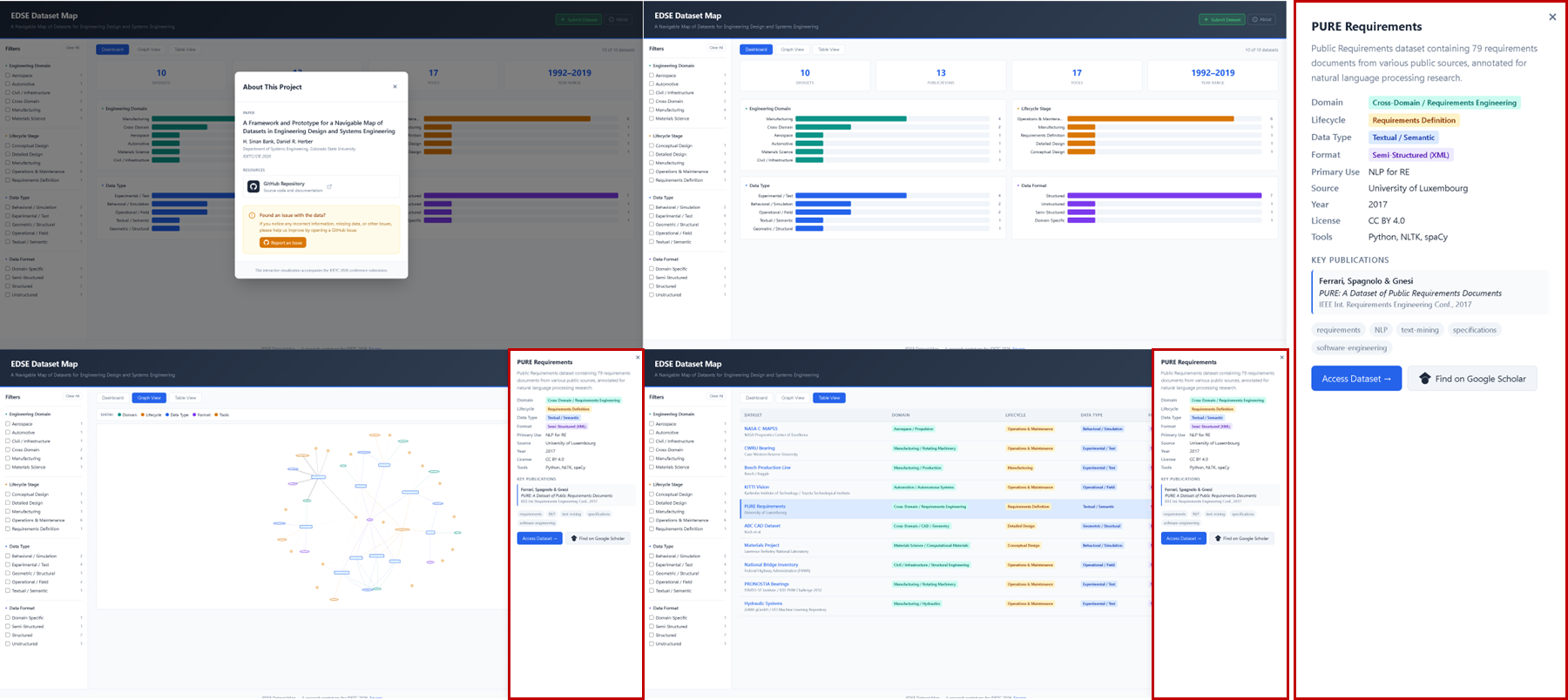}
\caption{Screenshots of the Map of Datasets in EDSE prototype: about modal (top left), dashboard (top right), knowledge graph view (bottom left), table view with filters (bottom right), and dataset detail panel (side).}
\label{fig:prototype}
\end{figure*}

\subsubsection{User Interaction and Querying.}~The user interface must support both exploratory browsing---for users who do not yet know what they are looking for---and directed querying---for users with specific requirements. Faceted search allows users to progressively filter datasets by selecting terms from any combination of the four taxonomic dimensions; each selection narrows the result set and updates the available filter options based on remaining datasets. Relationship navigation allows users to traverse the graph by following edges: starting from a publication, users can find all datasets it cites; starting from a tool, users can find compatible datasets; starting from a dataset, users can find all publications that have used it. Natural language querying allows users to express information needs conversationally (e.g., ``find me CAD datasets for aerospace components with open licenses''), with the system translating to structured graph queries.

\subsubsection{Visualization Paradigms.}~Effective visualization helps users understand the dataset landscape at a glance. A sunburst chart can visualize the hierarchical taxonomy structure, with ring segments representing taxonomy terms and segment size proportional to dataset counts, enabling users to immediately perceive where data is concentrated. A network graph can display relationships between datasets, publications, and tools as an interactive node-link diagram, revealing clusters of related resources and influential datasets that connect multiple communities. A lifecycle heatmap can cross-tabulate engineering domains against lifecycle stages, with cell color intensity indicating data availability, immediately revealing where data deserts and oases exist. A domain matrix can similarly cross-tabulate domains against data types or formats to identify underrepresented combinations.

\subsection{Dataset Cataloging Process}

To populate the map, a systematic cataloging process is employed. Identification involves discovering datasets through literature review, repository searches (e.g., Zenodo, Kaggle, NASA Open Data), competition archives (e.g., PHM Society Data Challenges), and community contributions. Profiling creates standardized metadata for each dataset, including source, description, access method, license, temporal and size characteristics, and known applications in the literature. Classification assigns datasets to terms along all four taxonomic dimensions, with multi-labeling where a dataset spans multiple categories. Relationship extraction identifies and encodes connections to publications that use the dataset, tools that process it, and related datasets (e.g., extended versions, derived subsets). Quality assessment records data quality indicators where available, including documentation completeness, FAIR compliance scores, and community feedback on usability.

%---------------
% Section 4: Results and Discussion
\xsection{Results and Discussion}\label{sec:results}

This section presents the results of applying the proposed framework, including an exemplar catalog of public datasets, a gap analysis of the current data landscape, and a discussion of synthetic data as a mitigation strategy.

\subsection{Exemplar Catalog of Public Datasets}

To illustrate the taxonomy's application, a curated catalog of diverse, publicly accessible datasets was compiled. Each entry was profiled with its source, description, classification across the four dimensions, and known applications. Table~\ref{tab:exemplar_datasets} presents three representative examples spanning different domains, lifecycle stages, and data types. 
% The potentially growing prototype of the Map of Datasets in its current state is available online (see Fig.~\ref{fig:prototype}).

\begin{table*}[htbp]
\def\arraystretch{1.2}
\centering
\caption{Representative Examples from the EDSE Dataset Catalog}
\label{tab:exemplar_datasets}
\small
% \begin{tabular}{p{3.5cm}p{2.5cm}p{2.5cm}p{2.5cm}p{2.5cm}p{2cm}}
\begin{tabular}{p{3.3cm}
>{\raggedright}p{2.1cm}
>{\raggedright}p{2.5cm}
>{\raggedright}p{3.1cm}
>{\raggedright}p{3.2cm}
p{2cm}
}
\hline \hline
\textbf{Dataset} & \textbf{Domain} & \textbf{Lifecycle Stage} & \textbf{Data Type} & \textbf{Format} & \textbf{Primary Use} \\
\hline
NASA C-MAPSS~\cite{Saxena2008} & Aerospace / Propulsion & Operations \& Maintenance & Behavioral / Simulation & Structured (CSV) & RUL Prediction \\
ABC CAD Dataset~\cite{ABC} & Cross-Domain & Detailed Design & Geometric / Structural & Domain-Specific (STEP) & Shape Analysis \\
PURE Requirements~\cite{Ferrari2017} & Cross-Domain & Requirements Definition & Textual / Semantic & Semi-Structured (XML) & NLP for RE \\
\hline \hline
\end{tabular}
\end{table*}

These three examples demonstrate the taxonomy's ability to classify datasets across contrasting dimensions---from operational simulation data to geometric models to textual requirements---enabling cross-cutting discovery.

\subsection{Interactive Discovery Tool Prototype}

To demonstrate the feasibility of the proposed framework, a web-based prototype was developed. The tool implements the multi-dimensional taxonomy as interactive faceted filters and provides multiple complementary views for dataset exploration. Figure~\ref{fig:prototype} presents screenshots of the prototype~\cite{EDSEMapApp2026}.

A dashboard view (Fig.~\ref{fig:prototype}, top right) provides an at-a-glance summary of the catalog, showing total datasets, publications, tools, and year range alongside distribution charts for each taxonomic dimension. The table view (Fig.~\ref{fig:prototype}, bottom right) displays datasets in a filterable, sortable table with taxonomy classifications rendered as color-coded badges. Selecting a dataset highlights the corresponding row and opens a detail panel (Fig.~\ref{fig:prototype}, side panel) with full metadata, including source, license, tools, key publications, and a Google Scholar search link. The graph view (Fig.~\ref{fig:prototype}, bottom left) renders the dataset catalog as an interactive knowledge graph using a force-directed layout. Dataset nodes (rectangles) connect to taxonomy term nodes (ellipses) colored by dimension---teal for domain, orange for lifecycle, blue for data type, and purple for format---with tool nodes (diamonds) shown in amber. Layer toggles allow users to show or hide specific dimensions to reduce visual complexity. An about modal (Fig.~\ref{fig:prototype}, top left) presents the project context and provides links for community contribution, including dataset submission and issue reporting through structured GitHub issue templates.

\subsection{Gap Analysis: Data Deserts and Data Oases}

Analysis of the current dataset landscape using the proposed taxonomy reveals significant imbalances in data availability across the engineering design space. The literature documents clear contrasts between underrepresented and well-supported areas.

\subsubsection{Data Deserts:~Underrepresented Areas.}~A significant scarcity of public datasets exists in specific domains and lifecycle stages. The literature explicitly documents these gaps:

\textbf{Early Lifecycle Stages:} Data related to conceptual design, requirements engineering, and trade studies are exceptionally rare. A prominent challenge facing machine learning adoption in engineering design research is the scarcity of publicly available, high-quality datasets~\cite{Ahmed2025}. The lack of readily accessible, specialized engineering design data---in contrast to fields such as computer vision---hinders research efforts.

\textbf{System Architecture and MBSE:} Publicly available system models (e.g., SysML repositories) are scarce. Progress in data-driven engineering design has been significantly slowed by the lack of standardized simulation environments and diverse datasets~\cite{Felten2025}.

\textbf{Late Lifecycle Stages:} The catalog reveals a near-complete absence of public datasets covering system disposal, retirement, and material recovery, suggesting these stages remain a critical data desert.

\textbf{Specific Domains:} Certain domains have limited open datasets due to proprietary, safety, or privacy concerns. A severe lack of datasets exists for mechanics, dynamics, and engineering design on platforms such as Hugging Face, contrasting sharply with natural language processing and computer vision~\cite{Ebel2025}.

The limited generalizability of current data-driven methods represents a major challenge, largely due to simplified models and constrained datasets~\cite{Afifi2025}. Research in testing, validation, and verification of autonomous systems reveals fragmentation in evaluation methodologies and a lack of consolidated benchmarks~\cite{Araujo2023}.

\subsubsection{Data Oases:~Well-Represented Areas.}
\begin{figure}[b]
\centering
\includegraphics[width=1\columnwidth]{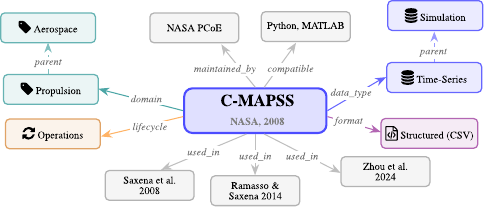}
\caption{Example of the C-MAPSS dataset represented in the proposed knowledge graph schema.}
\label{fig:kg_example}
\end{figure}

~Conversely, certain areas benefit from an abundance of high-quality public data, often propelled by community or institutional efforts, as illustrated in Fig.~\ref{fig:kg_example}:

%===============================================================================
% FIGURE: Knowledge Graph Example Instance
%===============================================================================

\textbf{Predictive Maintenance of Rotating Machinery:} A wealth of datasets (e.g., CWRU~\cite{CWRU}, PRONOSTIA~\cite{Nectoux2012}, NASA C-MAPSS~\cite{Saxena2008}) has created a robust benchmarking environment. Figure~\ref{fig:kg_example} illustrates how C-MAPSS is represented in the proposed knowledge graph schema, with taxonomy classification across all four dimensions and semantic connections to publications, tools, and the maintaining organization. The current prototype implements dataset, taxonomy term, and tool nodes; publication and organization nodes are planned for future versions. The lack of common datasets was an impediment to progress in prognostics, which motivated the generation of the C-MAPSS datasets~\cite{Ramasso2014a,Ramasso2014b}. Numerous publications have utilized C-MAPSS for data-driven algorithms, demonstrating significant research impact. The PHM Society and NASA continue to have a major impact on dataset provision~\cite{Hagmeyer2021}. Analysis of benchmarking datasets has established minimum sample sizes for effective data-driven modeling~\cite{Eker2012}, while standardized evaluation metrics tailored for prognostics enable fair algorithm comparison~\cite{Saxena2009}. Open-source tools such as \href{https://pypi.org/project/pyphm/}{PyPHM} assist researchers in accessing and preprocessing common industrial datasets to facilitate reproducibility~\cite{vonHahn2022,PyPHM}. C-MAPSS continues to be used for state-of-the-art aero-engine performance degradation models~\cite{Zhou2024,Fu2023}.

\textbf{Autonomous Driving and Vision:} Large-scale, multi-sensor datasets (e.g., KITTI, Waymo Open, nuScenes) have fueled rapid advancements in perception and planning algorithms.

\textbf{Materials Science:} The Materials Project~\cite{MaterialsProject} has made millions of computed material properties openly available, revolutionizing materials discovery workflows.

\textbf{Public Infrastructure:} Government-mandated datasets, such as the National Bridge Inventory~\cite{NBI}, have enabled large-scale statistical studies of infrastructure condition.

Having standard benchmark datasets enables researchers to reproduce and validate results, which is critical for data-driven method development~\cite{Soualhi2023}. Benchmark datasets such as C-MAPSS and XJTU-SY are crucial for enabling comparability and reproducibility~\cite{Sufi2025}.

\subsection{The Role of Synthetic Data}

For areas identified as data deserts, synthetic data generation offers a viable mitigation strategy. High-fidelity, physics-based simulations and generative AI models can produce datasets for training and testing new methods when real data is unavailable or proprietary.
However, challenges must be addressed:
\begin{itemize}[nosep,label=$\circ$]
\item \textbf{Reality Gap:} Models trained on synthetic data may fail to generalize to real-world conditions if the simulation does not capture all relevant physics and variability
\item \textbf{Validation:} The fidelity of synthetic data must be validated against real-world observations where possible
\item \textbf{Documentation:} Synthetic datasets must be clearly labeled and their generation process documented to enable proper interpretation\\
\end{itemize}

The proposed knowledge graph architecture can accommodate synthetic datasets by including metadata about generation methods, simulation tools used, and validation status, enabling users to make informed decisions about dataset suitability.

\subsection{Challenges and Mitigation Strategies}

The development and maintenance of a sustainable dataset map face several challenges:
\begin{itemize}[nosep,label=$\circ$]

\item \textbf{Data Quality and Standardization:} Datasets vary in quality and format. Mitigation strategies include enforcing a standardized metadata schema (e.g., DCAT), providing data quality metrics in dataset profiles, and encouraging community contributions for cleaning and annotation.

\item \textbf{Metadata and Discoverability:} Creating rich, consistent metadata is labor-intensive. This can be addressed through semi-automated curation, community-driven contributions with moderation workflows, and adoption of persistent identifiers such as DOIs.

\item \textbf{Intellectual Property and Accessibility:} Licenses must be clearly tracked and displayed. The map should prioritize verifiably open datasets and provide clear guidance for those with restricted access.

\item \textbf{Sustainability:} To prevent stagnation, the map must be a living resource. This requires a governance model that encourages community contributions, integration with publication workflows, and institutional support from professional societies or university consortia.

\end{itemize}

%---------------
% Section 5: Conclusion and Future Work
\xsection{Conclusion and Future Work}\label{sec:conclusion}

This paper has presented a concrete design for a ``Map of Datasets in EDSE'', addressing the critical challenge of data fragmentation that impedes data-driven research and practice.

\subsection{Summary of Contributions}

The contributions of this work include:
\begin{enumerate}
    \item A \textit{multi-dimensional taxonomy} with four dimensions (Engineering Domain, Lifecycle Stage, Data Type, and Data Format) that enables faceted classification and discovery of engineering datasets
    \item A \textit{knowledge graph architecture} for an interactive discovery tool, specifying node types, relationship types, and interaction modes that support rich semantic relationships between datasets, publications, tools, and taxonomy terms
    \item An \textit{exemplar catalog} demonstrating the taxonomy's application to diverse public datasets from prognostics, manufacturing, autonomous systems, and materials science
    \item A \textit{gap analysis} revealing significant ``data deserts'' in early-stage design and system architecture, contrasted with ``data oases'' in predictive maintenance and autonomous driving
    \item Identification of \textit{challenges and mitigation strategies} for sustainable curation, including the role of synthetic data for underrepresented areas
\end{enumerate}

The analysis confirms that data scarcity is a documented, widespread challenge limiting validation of data-driven methods in engineering design, while also demonstrating that coordinated community efforts (as in PHM) can create transformative benchmark ecosystems.

\subsection{Future Work: Toward an Intelligent, Agent-Driven Ecosystem}

Future work should focus on a complete implementation from the current prototype of this map and cultivating a community of contributors. The evolution of this platform is envisioned as an intelligent, agent-driven ecosystem.

Recent breakthroughs in AI, including self-supervised learning and geometric deep learning, are enabling new approaches to scientific discovery~\cite{Wang2023}. AI agents demonstrate capabilities including user-centered interaction, semantic knowledge extraction, intelligent reasoning, automation, and explainability~\cite{Anjelia2025}. AI is already used to automate archival workflows around capture and organization of collections~\cite{Colavizza2021}.

In a future state, the dataset map could leverage agentic AI in several ways:
\begin{itemize}[nosep,label=$\circ$]
    \item \textbf{Automated Discovery:} AI agents could proactively crawl repositories, publications, and institutional websites to identify new datasets, extracting metadata and proposing classifications
    \item \textbf{Semantic Enrichment:} Natural language processing could automatically extract relationships between datasets and publications, populating the knowledge graph with minimal manual effort
    \item \textbf{Quality Assessment:} Agents could evaluate dataset documentation completeness, FAIR compliance, and consistency, flagging issues for human review
    \item \textbf{Natural Language Querying:} Users could interact with the map using natural language queries (e.g., ``find me CAD datasets for aerospace components with permissive licenses''), with the agent translating to graph queries
\end{itemize}

% Data-centric AI focuses on enhancing data for building AI systems, including dataset discovery methods~\cite{Zha2025}. ``AI scientists'' are envisioned as collaborative agents integrating AI models with domain tools~\cite{Gao2024}. Current systems are entering the era of multimodal, agentic systems that can orchestrate software and hardware~\cite{Hartung2025}. Large repositories should enforce ontology-aligned data deposition for better machine interpretability.

% Scholarly knowledge graph construction techniques involve machine learning, rule-based learning, and NLP for automated curation and semantic enrichment~\cite{Verma2023}. The distinction between AI Agents (modular systems for task-specific automation) and Agentic AI (multi-agent collaboration with orchestrated autonomy) provides a conceptual framework for such systems~\cite{Sapkota2025}. AI is playing a transformative role in digital preservation within academic repositories, including automation of curation tasks via NLP for scalable metadata enrichment~\cite{Sousa2025}. 
% Platforms such as Sysrev demonstrate integration of machine learning with human review for data curation and systematic evidence review~\cite{Bozada2021}. These techniques can be adapted for engineering dataset catalogs.

\subsection{Call to Action}

By pursuing this vision, the Map of Datasets in EDSE can become an indispensable, self-sustaining infrastructure that empowers the next generation of data-driven engineering innovation. Success requires:
\begin{itemize}[nosep,label=$\circ$]

\item \textbf{Community Engagement:} Researchers, practitioners, and institutions contributing dataset entries and maintaining quality

\item \textbf{Institutional Support:} Professional societies (ASME, INCOSE) or university consortia providing governance and resources

\item \textbf{Integration with Workflows:} Linking dataset deposition to publication processes to ensure new datasets are discoverable

\item \textbf{Open Standards:} Adopting DCAT, DOIs, and FAIR principles to ensure interoperability with broader data ecosystems\\

\end{itemize}

The design presented here provides the foundation; realizing its potential requires coordinated action across the engineering design and systems engineering community.

%--------------------------

% \clearpage
\renewcommand{\refname}{REFERENCES}
\bibliographystyle{config/asmems4}
\begin{mySmall}
%\nocite{*} % remove later, displays all references
\bibliography{References}

@misc{EDSEMapApp2026,
  author       = {Bank, H. Sinan and Herber, Daniel R.},
  title        = {A Navigable Map of Datasets in Engineering Design and Systems Engineering: Interactive Web Application},
  year         = {2026},
  howpublished = {\url{https://map-of-edse-datasets.github.io/map-of-edse-datasets/}},
  note         = {Accessed: 2026},
}

@article{Fischer2020,
  title = {\href{https://doi.org/10.3390/make3010004}{{AI} System Engineering---Key Challenges and Lessons Learned}},
  author={Fischer, Lukas and Ehrlinger, Lisa and Geist, Verena and Ramler, Rudolf and Sobiezky, Florian and Zellinger, Werner and Brunner, David and Kumar, Mohit and Moser, Bernhard},
  journal={Machine Learning and Knowledge Extraction},
  volume={3},
  number={1},
  pages={56--83},
  year={2020},
  publisher={MDPI},
  url={https://doi.org/10.3390/make3010004}
}

@article{Briard2023,
  title = {\href{https://doi.org/10.1016/j.compind.2022.103814}{Challenges for Data-Driven Design in Early Physical Product Design: A Scientific and Industrial Perspective}},
  author={Briard, Tristan and Jean, Camille and Aoussat, Am{\'e}ziane and V{\'e}ron, Philippe},
  journal={Computers in Industry},
  volume={145},
  pages={103814},
  year={2023},
  publisher={Elsevier},
  url={https://doi.org/10.1016/j.compind.2022.103814}
}

@phdthesis{Rad2025,
  title = {\href{https://research.chalmers.se/publication/546108/file/546108_AdditionalFile_4a4a81dd.pdf}{Data Engineering for Data-Driven Design}},
  author={Rad, Mohammad Arjomandi},
  year={2025},
  url={https://research.chalmers.se/publication/546108/file/546108_AdditionalFile_4a4a81dd.pdf}
}

@inproceedings{Chami2022,
  title = {\href{https://doi.org/10.1002/iis2.12988}{Artificial Intelligence Capabilities for Effective Model-Based Systems Engineering: A Vision Paper}},
  author={Chami, Mohammad and Abdoun, Nabil and Bruel, Jean-Michel},
  booktitle={INCOSE International Symposium},
  volume={32},
  number={1},
  pages={1160--1174},
  year={2022},
  organization={Wiley Online Library},
  url={https://doi.org/10.1002/iis2.12988}
}

@article{Ahmed2025,
  title = {\href{https://doi.org/10.1115/1.4067871}{Special Issue: Design by Data: Cultivating Datasets for Engineering Design}},
  author={Ahmed, Faez and Picard, Cyril and Chen, Wei and McComb, Christopher and Wang, Pingfeng and Lee, Ikjin and Stankovic, Tino and Allaire, Douglas and Menzel, Stefan},
  journal={Journal of Mechanical Design},
  volume={147},
  number={4},
  pages={040301},
  year={2025},
  publisher={American Society of Mechanical Engineers},
  url={https://doi.org/10.1115/1.4067871}
}

@article{Felten2025,
  title = {\href{https://doi.org/10.48550/arXiv.2508.00831}{Engibench: A Framework for Data-Driven Engineering Design Research}},
  author={Felten, Florian and Apaza, Gabriel and Br{\"a}unlich, Gerhard and Diniz, Cashen and Dong, Xuliang and Drake, Arthur and Habibi, Milad and Hoffman, Nathaniel J and Keeler, Matthew and Massoudi, Soheyl and others},
  journal={arXiv preprint arXiv:2508.00831},
  year={2025},
  url={https://doi.org/10.48550/arXiv.2508.00831}
}

@article{Rudin2019,
  author = {Rudin, Cynthia},
  title = {\href{https://doi.org/10.1038/s42256-019-0048-x}{Stop Explaining Black Box Machine Learning Models for High Stakes Decisions and Use Interpretable Models Instead}},
  journal = {Nature Machine Intelligence},
  volume = {1},
  number = {5},
  pages = {206--215},
  year = {2019},
  publisher = {Springer Nature},
  url={https://doi.org/10.1038/s42256-019-0048-x}
}

@article{Schmidt2020,
  author = {Schmidt, Philipp and Biessmann, Felix and Teubner, Timm},
  title = {\href{https://doi.org/10.1080/12460125.2020.1819094}{Transparency and Trust in Artificial Intelligence Systems}},
  journal={Journal of Decision Systems},
  volume={29},
  number={4},
  pages={260--278},
  year={2020},
  publisher={Taylor \& Francis},
  url={https://doi.org/10.1080/12460125.2020.1819094}
}

@article{Wilkinson2016,
  title = {\href{https://doi.org/10.1038/sdata.2016.18}{The {FAIR} Guiding Principles for Scientific Data Management and Stewardship}},
  author={Wilkinson, Mark D and Dumontier, Michel and Aalbersberg, IJsbrand Jan and Appleton, Gabrielle and Axton, Myles and Baak, Arie and Blomberg, Niklas and Boiten, Jan-Willem and da Silva Santos, Luiz Bonino and Bourne, Philip E and others},
  journal={Scientific data},
  volume={3},
  number={1},
  pages={1--9},
  year={2016},
  publisher={Nature Publishing Group},
  url={https://doi.org/10.1038/sdata.2016.18}
}

@article{Ghose2024,
  title = {\href{https://doi.org/10.1007/s11367-024-02280-3}{Can {LCA} be {FAIR}? Assessing the Status Quo and Opportunities for {FAIR} Data Sharing}},
  author={Ghose, Agneta},
  journal={The International Journal of Life Cycle Assessment},
  volume={29},
  number={4},
  pages={733--744},
  year={2024},
  publisher={Springer},
  url={https://doi.org/10.1007/s11367-024-02280-3}
}

@inproceedings{Szykman2000,
  title = {\href{https://doi.org/10.1115/DETC2000/CIE-14622}{A Foundation for Interoperability in Next-Generation Product Development Systems}},
  author={Szykman, Simon and Fenves, Steven J and Keirouz, Walid and Shooter, Steven B},
  booktitle={International Design Engineering Technical Conferences and Computers and Information in Engineering Conference},
  volume={35111},
  pages={87--103},
  year={2000},
  organization={American Society of Mechanical Engineers},
  url={https://doi.org/10.1115/DETC2000/CIE-14622}
}

@article{Sadeghi2024,
  title = {\href{https://hdl.handle.net/11311/1261473}{Interoperability of Heterogeneous Systems of Systems: From Requirements to a Reference Architecture}},
  author={Sadeghi, Mersedeh and Carenini, Alessio and Corcho, Oscar and Rossi, Matteo and Santoro, Riccardo and Vogelsang, Andreas},
  journal={The Journal of Supercomputing},
  volume={80},
  number={7},
  pages={8954--8987},
  year={2024},
  publisher={Springer},
  url={https://hdl.handle.net/11311/1261473}
}

@article{Bodenbenner2021,
  title = {\href{https://doi.org/10.1016/j.measen.2021.100206}{{FAIR} Sensor Services---Towards Sustainable Sensor Data Management}},
  author={Bodenbenner, Matthias and Montavon, Benjamin and Schmitt, Robert H},
  journal={Measurement: Sensors},
  volume={18},
  pages={100206},
  year={2021},
  publisher={Elsevier},
  url={https://doi.org/10.1016/j.measen.2021.100206}
}

@article{Eickhoff2020,
  title = {\href{https://doi.org/10.1016/j.procir.2019.11.006}{A Metadata Repository for Semantic Product Lifecycle Management}},
  author={Eickhoff, Thomas and Eiden, Andreas and G{\"o}bel, Jens Christian and Eigner, Martin},
  journal={Procedia CIRP},
  volume={91},
  pages={249--254},
  year={2020},
  publisher={Elsevier},
  url={https://doi.org/10.1016/j.procir.2019.11.006}
}

@article{Hamlaoui2025,
  title = {\href{https://doi.org/10.1016/j.procir.2025.08.030}{Digital Twin for Field Data Management: Design of a Platform to Promote {FAIR} Principles and Ensure Data Reusability}},
  author={Hamlaoui, Rayen and Orimi, Atefeh Gooran and Donia, Reem and Backe, Christian and Briken, Veit and Lachmayer, Roland},
  journal={Procedia CIRP},
  volume={136},
  pages={165--170},
  year={2025},
  publisher={Elsevier},
  url={https://doi.org/10.1016/j.procir.2025.08.030}
}

@article{Karakoltzidis2024,
  title = {\href{https://doi.org/10.1039/D4SU00171K}{The {FAIR} Principles as a Key Enabler to Operationalize Safe and Sustainable by Design Approaches}},
  author={Karakoltzidis, Achilleas and Battistelli, Chiara Laura and Bossa, Cecilia and Bouman, Evert A and Aguirre, Irantzu Garmendia and Iavicoli, Ivo and Jeddi, Maryam Zare and Karakitsios, Spyros and Leso, Veruscka and L{\o}fstedt, Magnus and others},
  journal={RSC Sustainability},
  volume={2},
  number={11},
  pages={3464--3477},
  year={2024},
  publisher={Royal Society of Chemistry},
  url={https://doi.org/10.1039/D4SU00171K}
}

@article{Kitamura2006,
  title = {\href{https://www.designsociety.org/download-publication/25928/ROLES+OF+ONTOLOGIES+OF+ENGINEERING+ARTIFACTS+FOR+DESIGN+KNOWLEDGE+MODELING}{Roles of Ontologies of Engineering Artifacts for Design Knowledge Modeling}},
  author={Kitamura, Yoshinobu and others},
  year={2006},
  url={https://www.designsociety.org/download-publication/25928/ROLES+OF+ONTOLOGIES+OF+ENGINEERING+ARTIFACTS+FOR+DESIGN+KNOWLEDGE+MODELING}
}

@incollection{Borgo2009,
  title = {\href{https://doi.org/10.1016/B978-0-444-51667-1.50015-X}{Artefacts in Formal Ontology}},
  author={Borgo, Stefano and Vieu, Laure},
  booktitle={Philosophy of technology and engineering sciences},
  pages={273--307},
  year={2009},
  publisher={Elsevier},
  url={https://doi.org/10.1016/B978-0-444-51667-1.50015-X}
}

@article{Mario2013,
  title = {\href{https://doi.org/10.1007/978-3-642-11940-8}{Integrated Computer-Aided Design in Automotive Development}},
  author={Hirz, Mario and Dietrich, Wilhelm and Gfrerrer, Anton and Lang, Johann},
  journal={Springer-Verl. Berl.-Heidelb. DOI},
  volume={10},
  pages={978--3},
  year={2013},
  publisher={Springer},
  url={https://doi.org/10.1007/978-3-642-11940-8}
}

@article{Boussuge2019,
  title = {\href{https://doi.org/10.1080/09544828.2019.1630806}{Capturing Simulation Intent in an Ontology: {CAD} and {CAE} Integration Application}},
  author={Boussuge, Flavien and Tierney, Christopher M and Vilmart, Harold and Robinson, Trevor T and Armstrong, Cecil G and Nolan, Declan C and L{\'e}on, Jean-Claude and Ulliana, Federico},
  journal={Journal of Engineering Design},
  volume={30},
  number={10-12},
  pages={688--725},
  year={2019},
  publisher={Taylor \& Francis},
  url={https://doi.org/10.1080/09544828.2019.1630806}
}

@inproceedings{Zhu2009,
  title = {\href{https://doi.org/10.1115/DETC2009-87768}{Ontology-Driven Integration of {CAD}/{CAE} Applications: Strategies and Comparisons}},
  author={Zhu, Lijuan and Jayaram, Uma and Jayaram, Sankar and Kim, OkJoon},
  booktitle={International Design Engineering Technical Conferences and Computers and Information in Engineering Conference},
  volume={48999},
  pages={1461--1472},
  year={2009},
  url={https://doi.org/10.1115/DETC2009-87768}
}

@article{Zhan2010,
  title = {\href{https://doi.org/10.1115/1.3330432}{Knowledge Representation and Ontology Mapping Methods for Product Data in Engineering Applications}},
  author={Zhan, Pei and Jayaram, Uma and Kim, OkJoon and Zhu, Lijuan},
  year={2010},
  url={https://doi.org/10.1115/1.3330432}
}

@phdthesis{Zhan2007,
  title = {\href{https://doi.org/10.7273/000005683}{An Ontology-Based Approach for Semantic Level Information Exchange and Integration in Applications for Product Lifecycle Management}},
  author={Zhan, Pei},
  year={2007},
  school={Washington State University},
  url={https://doi.org/10.7273/000005683}
}

@article{Peachavanish2006,
  title = {\href{https://doi.org/10.1016/j.aei.2005.06.001}{An Ontological Engineering Approach for Integrating {CAD} and {GIS} in Support of Infrastructure Management}},
  author={Peachavanish, Ratchata and Karimi, Hassan A and Akinci, Burcu and Boukamp, Frank},
  journal={Advanced Engineering Informatics},
  volume={20},
  number={1},
  pages={71--88},
  year={2006},
  publisher={Elsevier},
  url={https://doi.org/10.1016/j.aei.2005.06.001}
}

@inproceedings{Summers2004,
  title = {\href{https://doi.org/10.1115/DETC2004-57514}{Representation in Engineering Design: A Framework for Classification}},
  author={Summers, Joshua D and Shah, Jami J},
  booktitle={International Design Engineering Technical Conferences and Computers and Information in Engineering Conference},
  volume={46962},
  pages={439--448},
  year={2004},
  url={https://doi.org/10.1115/DETC2004-57514}
}

@inproceedings{Witherell2013,
  title = {\href{https://doi.org/10.1115/IMECE2013-65220}{Towards the Synthesis of Product Knowledge Across the Lifecycle}},
  author={Witherell, Paul and Kulvatunyou, Boonserm and Rachuri, Sudarsan},
  booktitle={ASME International Mechanical Engineering Congress and Exposition},
  volume={56413},
  pages={V012T13A071},
  year={2013},
  organization={American Society of Mechanical Engineers},
  url={https://doi.org/10.1115/IMECE2013-65220}
}

@article{Hofer2024,
  title = {\href{https://doi.org/10.3390/info15080509}{Construction of Knowledge Graphs: Current State and Challenges}},
  author={Hofer, Marvin and Obraczka, Daniel and Saeedi, Alieh and K{\"o}pcke, Hanna and Rahm, Erhard},
  journal={Information},
  volume={15},
  number={8},
  pages={509},
  year={2024},
  publisher={MDPI},
  url={https://doi.org/10.3390/info15080509}
}

@phdthesis{DiPierro2025,
  title = {\href{https://hdl.handle.net/11586/539805}{Ontology-Enriched Graph Databases: An Interoperable Framework for Knowledge Integration and Management}},
  author={Di Pierro, Davide and others},
  year={2025},
  publisher={Universit{\`a} degli studi di Bari},
  url={https://hdl.handle.net/11586/539805}
}

@inproceedings{Rojas2021,
  title = {\href{https://doi.org/10.1007/978-3-030-88361-4_38}{Leveraging Semantic Technologies for Digital Interoperability in the European Railway Domain}},
  author={Rojas, Juli{\'a}n Andr{\'e}s and Aguado, Marina and Vasilopoulou, Polymnia and Velitchkov, Ivo and Van Assche, Dylan and Colpaert, Pieter and Verborgh, Ruben},
  booktitle={International Semantic Web Conference},
  pages={648--664},
  year={2021},
  organization={Springer},
  url={https://doi.org/10.1007/978-3-030-88361-4_38}
}

@article{Kalayci2021,
  title = {\href{https://doi.org/10.3390/su13031583}{A Knowledge Graph-Based Data Integration Framework Applied to Battery Data Management}},
  author={Kalayc{\i}, Tahir Emre and Bricelj, Bor and Lah, Marko and Pichler, Franz and Scharrer, Matthias K and Rube{\v{s}}a-Zrim, Jelena},
  journal={Sustainability},
  volume={13},
  number={3},
  pages={1583},
  year={2021},
  publisher={MDPI},
  url={https://doi.org/10.3390/su13031583}
}

@article{Wu2022,
  title = {\href{https://doi.org/10.1016/j.cageo.2022.105215}{{LinkClimate}: An Interoperable Knowledge Graph Platform for Climate Data}},
  author={Wu, Jiantao and Orlandi, Fabrizio and O’Sullivan, Declan and Dev, Soumyabrata},
  journal={Computers \& Geosciences},
  volume={169},
  pages={105215},
  year={2022},
  publisher={Elsevier},
  url={https://doi.org/10.1016/j.cageo.2022.105215}
}

@inproceedings{Zhang2025,
  title = {\href{https://doi.org/10.1007/978-981-96-3969-4_20}{Research on the Construction Method of Railway Data Resource Catalog Based on Knowledge Graphs}},
  author={Zhang, Kai and Sun, Siqi and Zou, Dan and Weng, Shengyuan and Ma, Xiaoning},
  booktitle={International Conference on Artificial Intelligence and Autonomous Transportation},
  pages={177--184},
  year={2024},
  organization={Springer},
  url={https://doi.org/10.1007/978-981-96-3969-4_20}
}

@inproceedings{Lorincz2020,
  title = {\href{https://doi.org/10.23919/MIPRO48935.2020.9245152}{Transforming Product Catalogue Relational into Graph Database: A Performance Comparison}},
  author={Lorincz, Josip and Huljic, Vlatka and Begusic, Dinko},
  booktitle={MIPRO},
  pages={523--528},
  year={2020},
  url={https://doi.org/10.23919/MIPRO48935.2020.9245152}
}

@inproceedings{Toledo2025,
  title = {\href{https://doi.org/10.1007/978-3-032-09530-5_23}{Using Semantic Technologies in the Railway Domain: The Register of Infrastructure ({RINF}) System}},
  author={Toledo, Jhon and Do{\~n}a, Daniel and Ruckhaus, Edna and Corcho, Oscar and Aguado, Marina and Patru, Dragos and Atemezing, Ghislain and Vasilopoulou, Polymnia},
  booktitle={International Semantic Web Conference},
  pages={398--414},
  year={2025},
  organization={Springer},
  url={https://doi.org/10.1007/978-3-032-09530-5_23}
}

@article{Saxena2008,
  title = {\href{https://www.nasa.gov/intelligent-systems-division/discovery-and-systems-health/pcoe/pcoe-data-set-repository/#:~:text=6.%20Turbofan%20Engine%20Degradation%20Simulation}{Turbofan Engine Degradation Simulation Data Set}},
  author={Saxena, Abhinav and Goebel, Kai},
  journal={NASA ames prognostics data repository},
  volume={18},
  pages={878--887},
  year={2008},
  url={https://www.nasa.gov/intelligent-systems-division/discovery-and-systems-health/pcoe/pcoe-data-set-repository/#:~:text=6.%20Turbofan%20Engine%20Degradation%20Simulation}
}

@misc{CWRU,
  author = {{Case Western Reserve University}},
  title = {\href{https://engineering.case.edu/bearingdatacenter}{Bearing Data Center}},
  year = {2000},
  url={https://engineering.case.edu/bearingdatacenter}
}

@inproceedings{Geiger2012,
  title = {\href{https://doi.org/10.1109/CVPR.2012.6248074}{Are We Ready for Autonomous Driving? {The} {KITTI} Vision Benchmark Suite}},
  author={Geiger, Andreas and Lenz, Philip and Urtasun, Raquel},
  booktitle={2012 IEEE conference on computer vision and pattern recognition},
  pages={3354--3361},
  year={2012},
  organization={IEEE},
  url={https://doi.org/10.1109/CVPR.2012.6248074}
}

@inproceedings{Ferrari2017,
  title = {\href{https://doi.org/10.1109/RE.2017.29}{{PURE}: A Dataset of Public Requirements Documents}},
  author={Ferrari, Alessio and Spagnolo, Giorgio Oronzo and Gnesi, Stefania},
  booktitle={2017 IEEE 25th international requirements engineering conference (RE)},
  pages={502--505},
  year={2017},
  organization={IEEE},
  url={https://doi.org/10.1109/RE.2017.29}
}

@inproceedings{ABC,
  title = {\href{https://doi.org/10.1109/CVPR.2019.00983}{{ABC}: A Big {CAD} Model Dataset For Geometric Deep Learning}},
  author={Koch, Sebastian and Matveev, Albert and Jiang, Zhongshi and Williams, Francis and Artemov, Alexey and Burnaev, Evgeny and Alexa, Marc and Zorin, Denis and Panozzo, Daniele},
  booktitle={Proceedings of the IEEE/CVF conference on computer vision and pattern recognition},
  pages={9601--9611},
  year={2019},
  url={https://doi.org/10.1109/CVPR.2019.00983}
}

@misc{MaterialsProject,
  author = {{Materials Project}},
  title = {Materials Project Database},
  howpublished = {\url{https://materialsproject.org}},
  year = {2011}
}

@misc{NBI,
  author = {{Federal Highway Administration}},
  title = {National Bridge Inventory},
  howpublished = {\url{https://www.fhwa.dot.gov/bridge/nbi.cfm}},
  year = {2023}
}

@inproceedings{Nectoux2012,
  title = {\href{https://hal.science/hal-00719503v1}{{PRONOSTIA}: An Experimental Platform for Bearings Accelerated Degradation Tests}},
  author={Nectoux, Patrick and Gouriveau, Rafael and Medjaher, Kamal and Ramasso, Emmanuel and Chebel-Morello, Brigitte and Zerhouni, Noureddine and Varnier, Christophe},
  booktitle={IEEE International Conference on Prognostics and Health Management, PHM'12.},
  pages={1--8},
  year={2012},
  organization={IEEE Catalog Number: CPF12PHM-CDR},
  url={https://hal.science/hal-00719503v1}
}

@article{Ebel2025,
  title = {\href{https://doi.org/10.1017/dce.2025.13}{Data Publishing in Mechanics and Dynamics: Challenges, Guidelines, and Examples from Engineering Design}},
  author={Ebel, Henrik and van Delden, Jan and L{\"u}ddecke, Timo and Borse, Aditya and Gulakala, Rutwik and Stoffel, Marcus and Yadav, Manish and Stender, Merten and Schindler, Leon and de Payrebrune, Kristin Miriam and others},
  journal={Data-Centric Engineering},
  volume={6},
  pages={e23},
  year={2025},
  publisher={Cambridge University Press},
  url={https://doi.org/10.1017/dce.2025.13}
}

@article{Afifi2025,
  title = {\href{https://doi.org/10.48550/arXiv.2511.20730}{Data-Driven Methods and {AI} in Engineering Design: A Systematic Literature Review Focusing on Challenges and Opportunities}},
  author={Afifi, Nehal and Wittig, Christoph and Paehler, Lukas and Lindenmann, Andreas and Wolter, Kai and Leitenberger, Felix and Dogru, Melih and Grauberger, Patric and D{\"u}ser, Tobias and Albers, Albert and others},
  journal={arXiv preprint arXiv:2511.20730},
  year={2025},
  url={https://doi.org/10.48550/arXiv.2511.20730}
}

@article{Ramasso2014a,
  title = {\href{https://hal.science/hal-01324587v1}{Performance Benchmarking and Analysis of Prognostic Methods for {CMAPSS} Datasets}},
  author={Ramasso, Emmanuel and Saxena, Abhinav},
  journal={International Journal of Prognostics and Health Management},
  volume={5},
  number={2},
  pages={1--15},
  year={2014},
  url={https://hal.science/hal-01324587v1}
}

@article{Hagmeyer2021,
  title = {\href{https://doi.org/10.36001/ijphm.2021.v12i2.3087}{Creation of Publicly Available Data Sets for Prognostics and Diagnostics Addressing Data Scenarios Relevant to Industrial Applications}},
  author={Hagmeyer, Simon and Mauthe, Fabian and Zeiler, Peter},
  journal={International Journal of Prognostics and Health Management},
  volume={12},
  number={2},
  year={2021},
  url={https://doi.org/10.36001/ijphm.2021.v12i2.3087}
}

@article{Soualhi2023,
  title = {\href{https://doi.org/10.36001/ijphm.2023.v14i2.3497}{Open Heterogeneous Data for Condition Monitoring of Multi Faults in Rotating Machines Used in Different Operating Conditions}},
  author={Soualhi, Moncef and Soualhi, Abdenour and Nguyen, Khanh TP and Medjaher, Kamal and Clerc, Guy and Razik, Hubert},
  journal={International Journal of Prognostics and Health Management},
  volume={14},
  number={2},
  year={2023},
  url={https://doi.org/10.36001/ijphm.2023.v14i2.3497}
}

@article{Sufi2025,
  title = {\href{https://doi.org/10.3390/app151910494}{Beyond the Sensor: A Systematic Review of {AI}'s Role in Next-Generation Machine Health Monitoring}},
  author={Sufi, Fahim},
  journal={Applied Sciences},
  volume={15},
  number={19},
  pages={10494},
  year={2025},
  publisher={MDPI},
  url={https://doi.org/10.3390/app151910494}
}

@inproceedings{Ramasso2014b,
  title = {\href{https://hal.science/hal-01145003v1}{Review and Analysis of Algorithmic Approaches Developed for Prognostics on {CMAPSS} Dataset}},
  author={Ramasso, Emmanuel and Saxena, Abhinav},
  booktitle={Annual Conference of the Prognostics and Health Management Society 2014.},
  year={2014},
  url={https://hal.science/hal-01145003v1}
}

@misc{PyPHM,
  title = {\href{https://pypi.org/project/pyphm/}{PyPHM: A Python Package for Prognostics and Health Management Datasets}},
  author={von Hahn, Tim},
  year={2022},
  note={Python package},
  url={https://pypi.org/project/pyphm/}
}

@article{vonHahn2022,
  title = {\href{https://doi.org/10.48550/arXiv.2205.15489}{Computational Reproducibility Within Prognostics and Health Management}},
  author={von Hahn, Tim and Mechefske, Chris K},
  journal={arXiv preprint arXiv:2205.15489},
  year={2022},
  url={https://doi.org/10.48550/arXiv.2205.15489}
}

@article{Fu2023,
  title = {\href{https://doi.org/10.3390/s23198124}{Prognostic and Health Management of Critical Aircraft Systems and Components: An Overview}},
  author={Fu, Shuai and Avdelidis, Nicolas P},
  journal={Sensors},
  volume={23},
  number={19},
  pages={8124},
  year={2023},
  publisher={MDPI},
  url={https://doi.org/10.3390/s23198124}
}

@inproceedings{Eker2012,
  title = {\href{https://doi.org/10.36001/phme.2012.v1i1.1409}{Major Challenges in Prognostics: Study on Benchmarking Prognostics Datasets}},
  author={Eker, Omer Faruk and Camci, Faith and Jennions, Ian K},
  booktitle={Phm society european conference},
  volume={1},
  number={1},
  year={2012},
  url={https://doi.org/10.36001/phme.2012.v1i1.1409}
}

@inproceedings{Saxena2009,
  title = {\href{https://doi.org/10.1109/AERO.2009.4839666}{Evaluating Algorithm Performance Metrics Tailored for Prognostics}},
  author={Saxena, Abhinav and Celaya, Jos{\'e} and Saha, Bhaskar and Saha, Sankalita and Goebel, Kai},
  booktitle={2009 IEEE Aerospace conference},
  pages={1--13},
  year={2009},
  organization={IEEE},
  url={https://doi.org/10.1109/AERO.2009.4839666}
}

@article{Zhou2024,
  title = {\href{https://doi.org/10.1109/ACCESS.2024.3460872}{Data-Driven Modeling of Aero-Engine Performance Degradation Models}},
  author={Zhou, Mingyang and Miao, Keqiang and Sun, Jiaxian and Shen, Yafeng and Han, Bo},
  journal={IEEE Access},
  volume={12},
  pages={150020--150031},
  year={2024},
  publisher={IEEE},
  url={https://doi.org/10.1109/ACCESS.2024.3460872}
}

@article{Araujo2023,
  title = {\href{https://doi.org/10.1145/3542945}{Testing, Validation, and Verification of Robotic and Autonomous Systems: A Systematic Review}},
  author={Araujo, Hugo and Mousavi, Mohammad Reza and Varshosaz, Mahsa},
  journal={ACM Transactions on Software Engineering and Methodology},
  volume={32},
  number={2},
  pages={1--61},
  year={2023},
  publisher={ACM New York, NY},
  url={https://doi.org/10.1145/3542945}
}

@article{Wang2023,
  title = {\href{https://doi.org/10.1038/s41586-023-06221-2}{Scientific Discovery in the Age of Artificial Intelligence}},
  author={Wang, Hanchen and Fu, Tianfan and Du, Yuanqi and Gao, Wenhao and Huang, Kexin and Liu, Ziming and Chandak, Payal and Liu, Shengchao and Van Katwyk, Peter and Deac, Andreea and others},
  journal={Nature},
  volume={620},
  number={7972},
  pages={47--60},
  year={2023},
  publisher={Nature Publishing Group UK London},
  url={https://doi.org/10.1038/s41586-023-06221-2}
}

@article{Anjelia2025,
  title = {\href{https://doi.org/10.59395/ijadis.v6i3.1462}{{AI} Agents for Organizational Knowledge Retrieval and Sharing: A Systematic Literature Review}},
  author={Anjelia, Sri Rosa and Sensuse, Dana Indra and Lusa, Sofian},
  journal={International Journal of Advances in Data and Information Systems},
  volume={6},
  number={3},
  pages={824--839},
  year={2025},
  url={https://doi.org/10.59395/ijadis.v6i3.1462}
}

@article{Colavizza2021,
  title = {\href{https://doi.org/10.1145/3479010}{Archives and {AI}: An Overview of Current Debates and Future Perspectives}},
  author={Colavizza, Giovanni and Blanke, Tobias and Jeurgens, Charles and Noordegraaf, Julia},
  journal={ACM Journal on Computing and Cultural Heritage (JOCCH)},
  volume={15},
  number={1},
  pages={1--15},
  year={2021},
  publisher={ACM New York, NY},
  url={https://doi.org/10.1145/3479010}
}
\end{mySmall}

%--------------------------
% \clearpage
% \onecolumn
% \xneed{Will be removed in the final version.}
% % This will be removed in the final version.
% \tableofcontents
% \listoffigures
% \listoftables
% -----------

%---------------
%  \clearpage
% \input{input/working.tex}

\end{document}